# One million percent tunnel magnetoresistance in a magnetic van der Waals heterostructure

Hyun Ho Kim[1,2], Bowen Yang[1,3], Tarun Patel[1,3], Francois Sfigakis[1,2], Chenghe Li[4], Shangjie Tian[4], Hechang Lei[4], and Adam W. Tsen[1,2*]

[1]*Institute for Quantum Computing, University of Waterloo, Waterloo, Ontario N2L 3G1, Canada*
[2]*Department of Chemistry, University of Waterloo, Waterloo, Ontario N2L 3G1, Canada*
[3]*Department of Physics and Astronomy, University of Waterloo, Waterloo, Ontario N2L 3G1, Canada*
[4]*Department of Physics and Beijing Key Laboratory of Opto-electronic Functional Materials & Micro-Nano Devices, Renmin University of China, Beijing 100872, China*

[*]Correspondence to: awtsen@uwaterloo.ca

## Abstract

We report the observation of a very large negative magnetoresistance effect in a van der Waals tunnel junction incorporating a thin magnetic semiconductor, $CrI_3$, as the active layer. At constant voltage bias, current increases by nearly one million percent upon application of a 2 Tesla field. The effect arises from a change between antiparallel to parallel alignment of spins across the different $CrI_3$ layers. Our results elucidate the nature of the magnetic state in ultrathin $CrI_3$ and present new opportunities for spintronics based on two-dimensional materials.

## Main Text

The exploration of two-dimensional (2D) materials beyond graphene is experiencing a new surge. In addition to semiconducting transition metal dichalcogenides,[1] compounds exhibiting collective electron phases, such as charge density waves,[2-9] superconductivity,[10-15] and magnetism,[16-23] have recently been stabilized down to single or few atomic layers. Not only are the 2D states different from their bulk counterparts, the incorporation of such materials into novel heterostructures will allow for the development of multifunctional electronic devices.[24, 25] Here, we demonstrate a tunnel junction heterostructure consisting of magnetic semiconductor, $CrI_3$, sandwiched between few-layer graphene electrodes and sealed with hexagonal boron nitride (hBN). Under voltage biasing, we observe a negative magnetoresistance close to one million percent at low temperature under an applied field of ~2T, a value on par with that of colossal magnetoresistance (CMR) in the manganites.[26, 27] We attribute the cause to a spin filtering effect arising from the full polarization of spins in the magnetic semiconductor,[28, 29] and experimentally determine the exchange gap through magnetotransport measurements.

In Fig. 1a, we show a schematic illustration of the device. Within a nitrogen-filled glovebox, thin $CrI_3$ was electrically contacted above and below by few-layer graphene and fully encapsulated with hBN to prevent reactions with the ambient environment. We confirmed that our $CrI_3$ bulk crystal has similar properties with previous reports (see Supplementary Materials, Fig. S1 and S2).[17, 30] The assembly and fabrication procedure is described in greater detail in Supplementary Materials, where we also show an optical image of the finished device (see Supplementary Materials, Fig. S3). Since $CrI_3$ is an intrinsic semiconductor,[31, 32] we expect conduction to occur primarily via tunneling in thin samples. We have measured the resistance-area product at room temperature and low bias ($V_{AC}$ = 4mV) for several such devices consisting of different $CrI_3$ thicknesses and the results are plotted in Fig. 1b. The resistance initially increases exponentially with thickness, indicative of quantum tunneling across the $CrI_3$ layers. Most of the following data were taken from the 10-layer-thick sample, although similar features were also seen in thicker samples. The temperature dependence of the low-bias resistance follows Arrhenius behavior down to ~160K, below which the current becomes too small to measure with our instruments (see Supplementary Materials, Fig. S4). In Fig. 1c, we show full current-voltage characteristics as a function of temperature. Current increases nonlinearly with increasing bias, as expected for a tunnel junction.[33] Overall, the effect of increasing temperature is to lower the voltage needed to achieve a certain current level.

The current rises dramatically under a magnetic field at low temperature. In the main panel of Fig. 1d, we show current as a function of field applied both perpendicular and parallel to the CrI$_3$ layers. For the former, current rises in discrete steps and at ~2T saturates to a value over four orders of magnitude larger than that at zero field, while for the latter it increases continuously without saturation up to 5.5T. This change is remarkable and is the central finding of this work. Defining the magnetoresistance percentage at constant voltage to be MR (%) = $\frac{I(B)-I_{min}}{I_{min}} \times$ 100%,[29] we find that the effect is four orders of magnitude larger than giant magnetoresistance (GMR) observed in Co/Cu magnetic multilayers[34] and on par with CMR in the mixed-valence manganites.[23, 24] See Supplementary Materials, Fig. S5 for a discussion of the measurement of $I_{min}$.

In the inset of Fig. 1d, we show the dependence of MR on applied voltage. We see that over a broad range, the MR exceeds $10^4$%. In a narrower voltage range, the MR almost reaches $10^6$%. Additional information describing the non-monotonic bias dependence as well as reproducible MR behavior in other CrI$_3$ devices can be found in Supplementary Materials, Fig. S6. There is also noticeable asymmetry between positive and negative voltage. While we are unclear as to the nature of this effect, we note that there is an inherent asymmetry built into the device geometry as the top and bottom graphite layers are not identical.

Our remaining discussion will be focused on understanding the origin of this very large magnetoresistance effect as well as its implications on the nature of magnetism in 2D CrI$_3$. We first measured the current-voltage dependence under different magnetic fields at low temperature, and the results are plotted in the main panels of Fig. 2a and 2b for perpendicular and parallel field configurations, respectively. We observe that, similar to increasing temperature, in both cases increasing field lowers the voltage necessary to achieve a given current level. The large rise in current thus results from this shift in conjunction with the highly nonlinear current-voltage characteristics of the junction. The difference between the two field orientations are captured in the insets, which show the measured voltage at 100nA current as a function of field intensity. The voltage decreases discretely (continuously) for perpendicular (parallel) field, consistent with the results of Fig. 1d. We note that due to the device nonlinearity, the very large MR values can only be achieved under constant voltage (and not constant current) biasing conditions.

We have performed similar measurements at different temperatures and the results are shown in Fig. 2c (perpendicular field) and 2d (parallel field). At zero field, the temperature-dependent voltage measured for 100nA current exhibits a marked kink at $T^*$ ~ 46K, below which it rises more steeply. We have observed this kink in other devices as well consisting of 14 and 20 $CrI_3$ layers—the kink temperature increases slightly with thickness (see Supplementary Materials, Fig. S7). As will be discussed later, we assign this temperature to be the onset of in-plane ferromagnetism in thin $CrI_3$, above which the sample behaves as an isotropic spin paramagnet. For comparison, we also show temperature-dependent magnetization taken for the bulk crystal in Fig. S2, which shows a slightly higher magnetic transition temperature: $T_c^{bulk}$ ~ 60K. [17, 30] For the junction device, applying field suppresses the voltage, with the largest effect seen at low temperature. This change gradually disappears above $T^*$ for either field direction. The very large negative magnetoresistance observed is thus intimately tied to the formation of the 2D magnetic state.

The key to understanding this phase lies in the field anisotropy of magnetoresistance. In the main panels of Fig. 3a and 3b, we show slow (0.2T/min) field sweeps of voltage taken at different temperatures for perpendicular and parallel field orientations, respectively. At low temperature, the former shows a series of abrupt jumps at low field as well as marked hysteresis between different sweep directions. The effect of hysteresis is highlighted in the inset of Fig. 3a, which shows a zoom-in of the 1.4K data around zero field. We also observed similar effects in other samples (see Supplementary Materials, Fig. S8 and S9), and the jumps in 14-layer $CrI_3$ device are highlighted in the main panel of Fig. 3c. Above ~2T, the voltage saturates to a field independent value of ~0.5V. As the temperature is increased above $T^*$, both the hysteresis and the jumps disappear, while the magnetoresistance decreases substantially. For a given temperature, the overall magnetoresistance change for parallel field is qualitatively similar, except hysteresis and abrupt jumps are never observed. Furthermore, at low temperature, the necessary field in order to achieve the same voltage change is significantly higher (~5.5T for ~0.5V at 1.4K). Interestingly, the two voltage curves become nearly identical at high temperature, as evident from the left panel of Fig. 3d. In the right panel, we have plotted the temperature-dependent magnetoresistance anisotropy, defined by: $\frac{V(B_\parallel) - V(B_\perp)}{[V(B_\parallel) + V(B_\perp)]/2} \times 100\%$, evaluated at 100nA current and 2.5T field. The anisotropy is nearly zero above $T^*$ and increases considerably below. Taken together, the results from Fig. 3 indicate that thin $CrI_3$ behaves as an isotropic paramagnet above this critical temperature.

The hysteresis and jumps observed below $T^*$ for perpendicular field are suggestive of domain flips analogous to the Barkhausen effect.[35] Indeed, magneto-optical measurements by Zhong et al. on $CrI_3$ flakes of comparable thickness report complex domain dynamics within a similar field range.[20] One possibility is that the different layers are coupled ferromagnetically and all flip spin within a given domain. In this case, lateral transport may be very sensitive to spin flip scattering at the domain walls. Since we measure vertical transport across the $CrI_3$ layers, however, we expect magnetoresistance from this change to be small. Alternatively, it is possible for the spins to flip layer by layer.[36] The resistance of such as state would thus depend sensitively on the relative spin orientation of adjacent layers.[29, 34] We have measured the number of jumps for junctions with different $CrI_3$ thickness, and we observe a larger number of jumps for thicker samples (see Fig. 3c, inset), which would further substantiate the latter scenario.

We thus propose the following model to account for the very large tunnel magnetoresistance observed in our devices at low temperature, which is shown schematically in Fig. 4a-c. In the paramagnetic phase above $T^*$, carriers encounter a spin-degenerate tunnel barrier, $\Phi_{PM}$ (dotted black line in Fig. 4c), from semiconducting $CrI_3$. Below $T^*$, the spins on each layer $CrI_3$ layer spontaneously polarize along the out-of-plane (easy axis) direction. In contrast with bulk crystals, thin $CrI_3$ exhibits antiferromagnetic (AFM) coupling between layers in the absence of field, yielding net zero moment. Such a state has been observed previously for bilayer samples.[17, 18] The result is a spin modulated tunnel barrier, whose relative large, effective barrier height, $\Phi_{AFM}$, is marked by the dotted blue line in Fig. 4c. Under a perpendicular field, the spins flip layer by layer, causing jumps in the magnetoresistance until an interlayer ferromagnetic (FM) state is reached at ~2T. We note that this field is larger than the saturation field in bulk $CrI_3$, possibility due to different interlayer magnetic interactions in the two systems. The FM state has a conduction band that is spin-split by the exchange interaction uniformly across all the layers. This spin filtering state has the smallest tunnel barrier, $\Phi_{FM}$, compared to both the PM and AFM barriers, and so also shows the smallest resistance. It shares similar properties with FM tunnel barriers EuO and EuS under zero-field conditions.[28, 37] Under an in-plane field, the anti-parallel spins in the AFM state cant continuously until a fully in-plane polarized state is reached at ~5.5T with similarly low resistance.

The barrier heights as well as the magnitude of the exchange gap between up and down spins can be approximated by fitting to transport described by Fowler-Nordheim tunneling.[37, 38] In this regime, current is given by:

$$J = \left(\frac{e^3 E^2}{4\hbar\Phi}\right) \exp\left(-\frac{4\sqrt{2m\Phi^3}}{3\hbar eE}\right)$$

where $e$ is electronic charge, $E$ is the electric field across the CrI$_3$, $m$ is the effective electron mass, which we estimate to be the free-electron mass, $\hbar$ is the Planck's constant, and $\Phi$ is the barrier height between the few-layer graphene injector and CrI$_3$ (averaged over several layers). We have plotted $\ln\left(\frac{I}{V^2}\right)$ vs. $V^{-1}$ in Fig. 4d and extracted $\Phi$ from the negative linear slope (marked by dashed red line) for the AFM state at zero field (1.4K), PM state at 49K (0T), and FM state at B ⊥ ab = 2.5T (1.4K). $\Phi_{AFM}$ ($\Phi_{FM}$) is larger (smaller) than $\Phi_{PM}$. We then estimate the exchange splitting as $E_{ex} = 2(\Phi_{PM} - \Phi_{FM})$.[39] In Fig. 4e, we plotted the different barrier heights and $E_{ex}$ for three devices of different CrI$_3$ thickness. The deduced splitting is comparable to the spin gap value obtained from bandstructure calculations for the conduction band,[27-29] and shows a small decrease with decreasing number of layers, which could explain why $T^*$ is slightly reduced from $T_c^{bulk}$. We have also measured $\Phi_{FM}$ for several different temperatures, which allows the exchange gap to be determined as a function of temperature. As shown in Fig. 4f, $E_{ex}$ decreases sharply when the temperature approaches $T^*$.

This work opens up new opportunities for spintronics incorporating magnetic 2D materials.

Note:

We became aware of several related works after the completion of our work.[40-42]

## Acknowledgements

We thank Guo-Xing Miao, Junichi Okamoto, and Liuyan Zhao for useful discussions. AWT acknowledges support from an NSERC Discovery grant (RGPIN-2017-03815) and the Korea - Canada Cooperation Program through the National Research Foundation of Korea (NRF) funded by the Ministry of Science, ICT and Future Planning (NRF-2017K1A3A1A12073407). This research was undertaken thanks in part to funding from the Canada First Research Excellence Fund, the Ministry of Science and Technology of China (No. 2016YFA0300504) and the National Natural Science Foundation of China (Grant No. 11574394, 11774423).

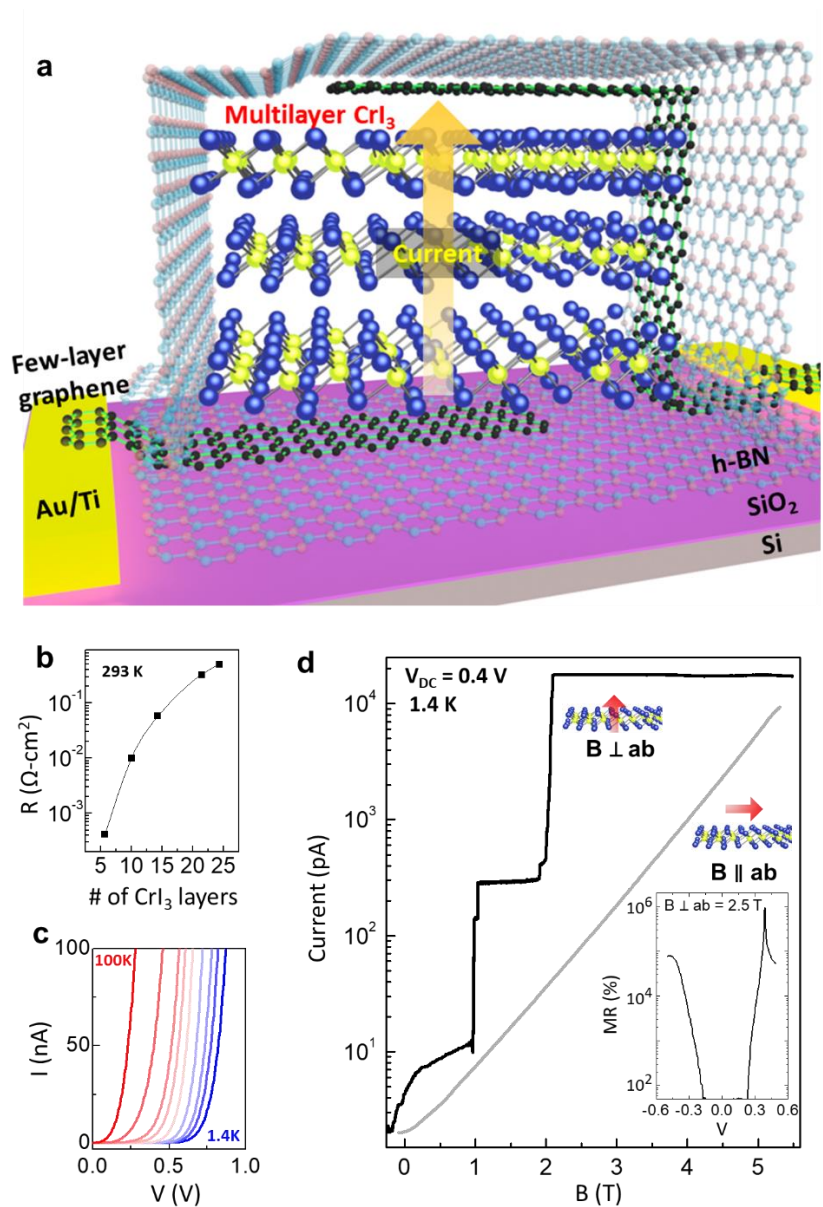

**Figure 1.** Vertical van der Waals heterojunction device incorporating magnetic $CrI_3$. (a) Schematic illustration of the device. (b) Area-normalized junction resistance vs. $CrI_3$ thickness at 293K. (c) Current vs. voltage at different temperatures (100, 80, 70, 60, 50, 40, 30, 20, and 1.4K in sequence from left) of a 10-layer $CrI_3$ junction. (d) Current vs. magnetic field at 0.4V. Inset shows voltage-dependent MR at 2.5T for the same device.

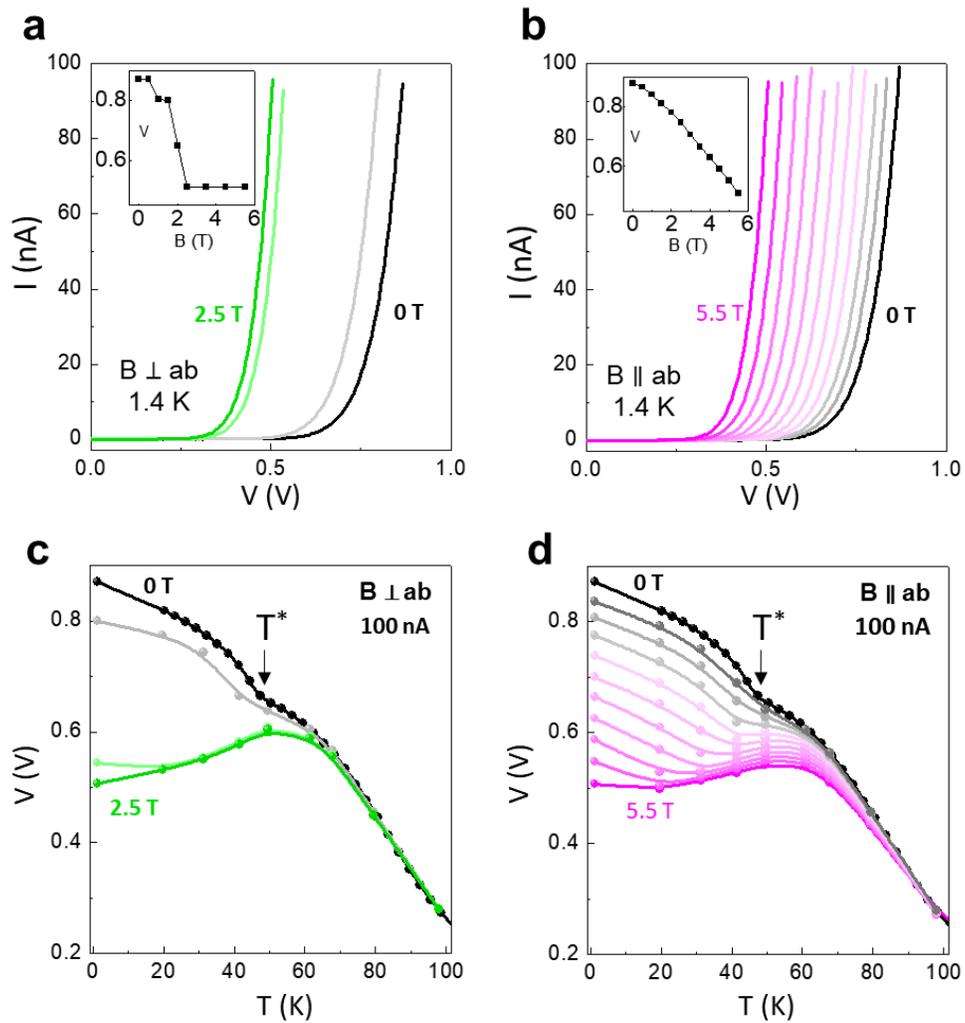

**Figure 2.** Field- and temperature-dependent transport measurements. (a) Current vs. voltage at 1.4K for different B⊥ab (0, 1, 2, and 2.5T, in sequence from right) (b) Same for B ∥ ab (from right: 0T, 1T to 5.5T with 0.5T steps). Insets show voltage vs. B for the two field orientations at 100nA. (c) Temperature-dependent voltage (left axis) for same B⊥ab above. (d) Same for B ∥ ab.

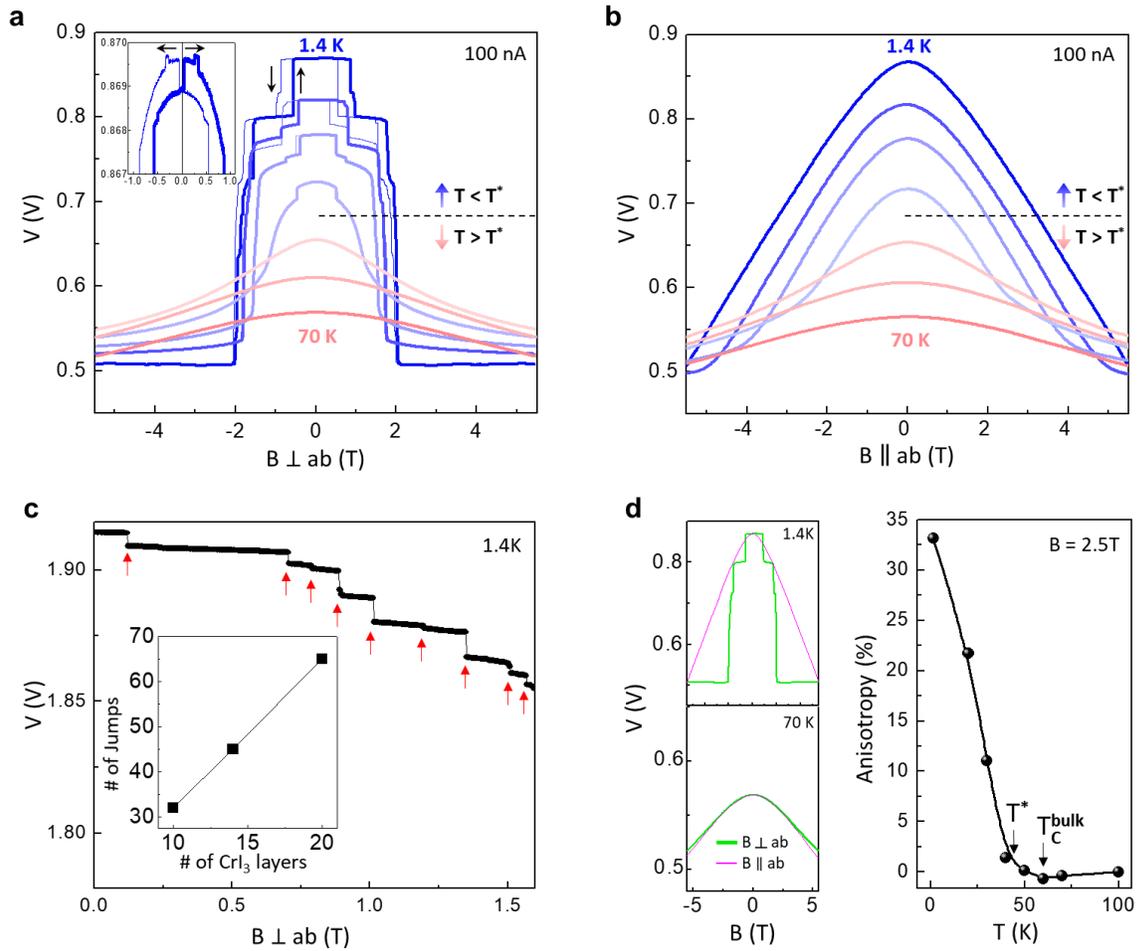

**Figure 3.** Magnetoresistance anisotropy and domain flipping in thin $CrI_3$. (a) Voltage vs. $B \perp ab$ sweep for 10-layer $CrI_3$ device at different temperatures (1.4, 20, 30, 40, 50, 60, and 70K, in sequence from top); (b) Same for $B \parallel ab$. Inset in (a) shows zoom-in of 1.4K data. (c) V vs. $B \perp ab$ showing many abrupt jumps for 14-layer $CrI_3$ device indicated by red arrows. Insets shows the total number of jumps vs. the number of $CrI_3$ layers. (d) Left: comparison of V vs. B sweeps at 100nA for two field orientations at two different temperature conditions (1.4 and 70 K). Right: magnetoresistance anisotropy at 2.5T vs. temperature.

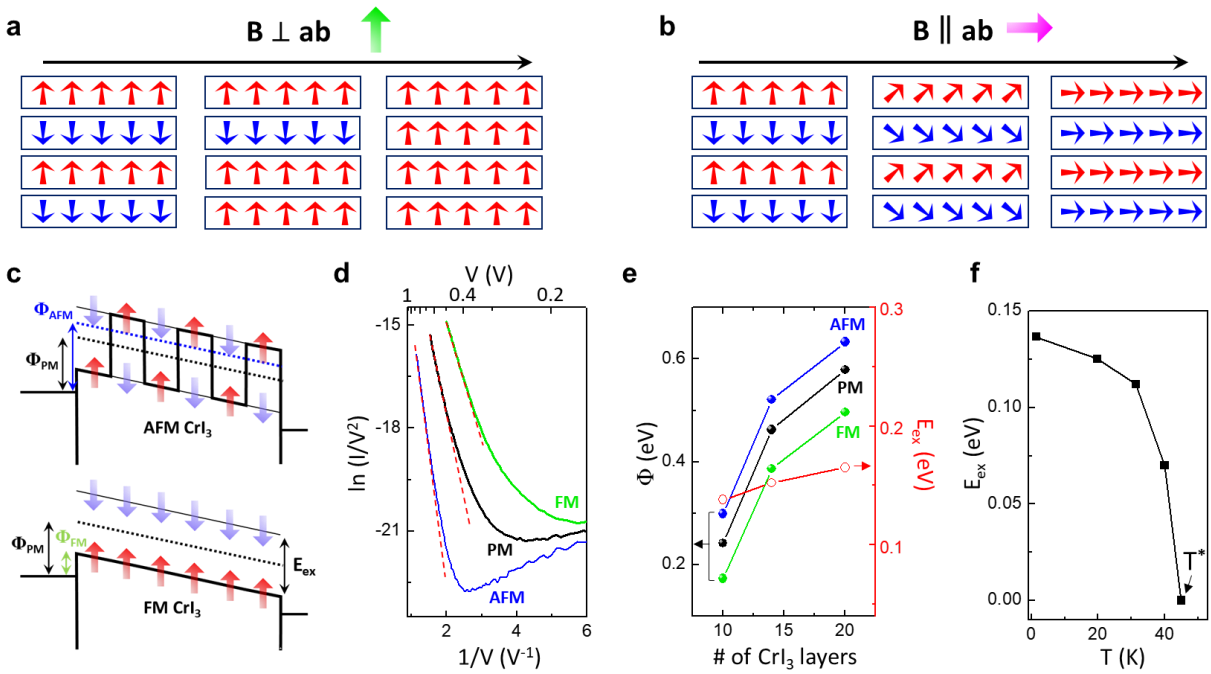

**Figure 4.** Interlayer spin coupling and origin of very large negative magnetoresistance. (a) Proposed transition mechanism from interlayer AFM to FM spin state with B⊥ab and (b) B∥ab. (c) Schematic energy diagrams with AFM barrier (top) and FM barrier (bottom) under Fowler-Nordheim tunneling. PM barrier is shown by black dashed line. (d) ln (I/V$^2$) versus V$^{-1}$ plot in three magnetic states and slope (marked by dashed red line) used to extract the height of the tunnel barrier. AFM, PM, and FM states were measured at 0T at 1.4K, 0T at 49K, and B⊥ab = 2.5T at 1.4K, respectively. (e) Thickness dependence of AFM, FM, and PM barrier height values (left axis) and deduced exchange splitting gap (right axis). (f) Exchange gap of 10-layer CrI$_3$ vs. temperature.

# Supplementary Materials for

## One million percent tunnel magnetoresistance in a magnetic van der Waals heterostructure

Hyun Ho Kim, Bowen Yang, Tarun Patel, Francois Sfigakis, Chenghe Li, Shangjie Tian, Hechang Lei, and Adam W. Tsen[*]

correspondence to: awtsen@uwaterloo.ca

**This PDF file includes:**

Materials and Methods
Bulk $CrI_3$ characterization
Optical image of $CrI_3$ device
Temperature-dependent AC conductance of $CrI_3$ device
Measurement of $I_{min}$
$CrI_3$ thickness dependence on (magneto)transport
Additional magnetic field sweep data
Figs. S1 to S9

## 1. Materials and Methods

Crystal synthesis.

Single crystals of CrI$_3$ were grown by chemical vapor transport method. Chromium powder (99.99%) and iodine flake (99.999%) in 1:3 molar ratio were put into the silica tube with the length of 200 mm, and inner diameter of 14 mm. The tube was evacuated down to 0.01 Pa and sealed under vacuum. The tubes were placed in two-zone horizontal tube furnace and the source and growth zones were raised to 923 K and 823 K for 3 days, and then held there for 7 days. The shiny, black, and plate-like crystals with lateral dimensions up to several millimeters can be obtained. In order to avoid degradation of CrI$_3$ crystals, the samples were stored in glovebox.

Device fabrication.

We have exfoliated few-layer h-BN (HQ graphene), graphene (CoorsTek), and CrI$_3$ on polydimethylsiloxane- (PDMS) based gel (PF-40/17-X4 from Gel-Pak) within a nitrogen-filled glove box (Inert Pure LabHE, $P_{O_2}$, $P_{H_2O}$ < 0.1 ppm). We fabricated Au (40 nm)/Ti (5nm) electrodes on 285 nm-thick SiO$_2$/Si, which were patterned by conventional photolithography and lift-off method with an e-beam deposition. A home-built transfer setup inside the glove box was used to stack vertical heterostructures of h-BN/few-layer graphene/CrI$_3$/few-layer graphene/h-BN onto the pre-patterned electrodes. Few-layer graphene flakes were contacted with the pre-patterned Au/Ti electrodes while h-BN provided a dielectric oxidation barrier. We sequentially transferred the flakes at room temperature by avoiding the bubbles and cracks generated during transfer process. After stacking, devices were annealed at 393 K in the glove box and were stored in vacuum desiccator until the devices were loaded into a cryostat.

## 2. Bulk crystal characterization

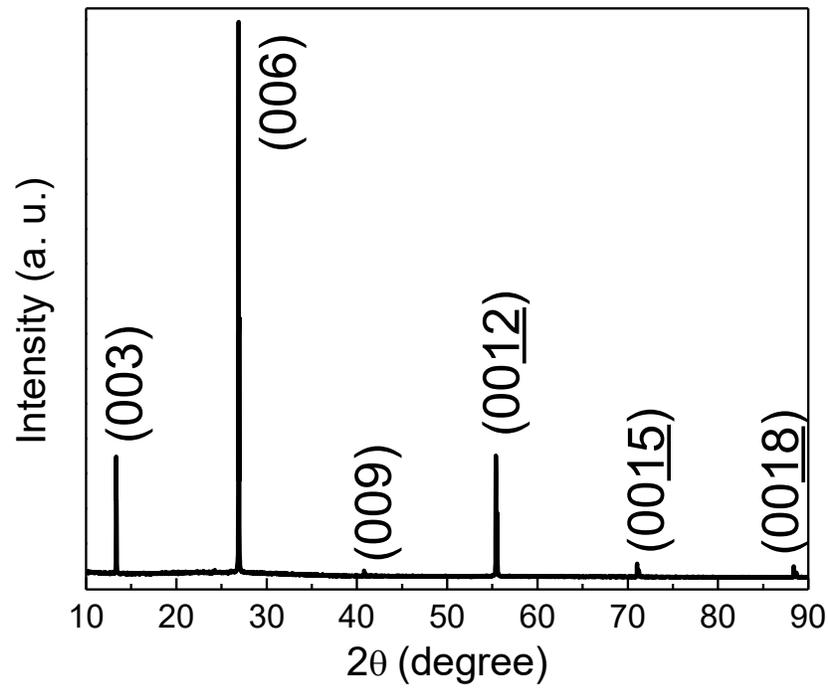

**Fig. S1**
X-ray diffraction pattern of bulk CrI$_3$ crystal.

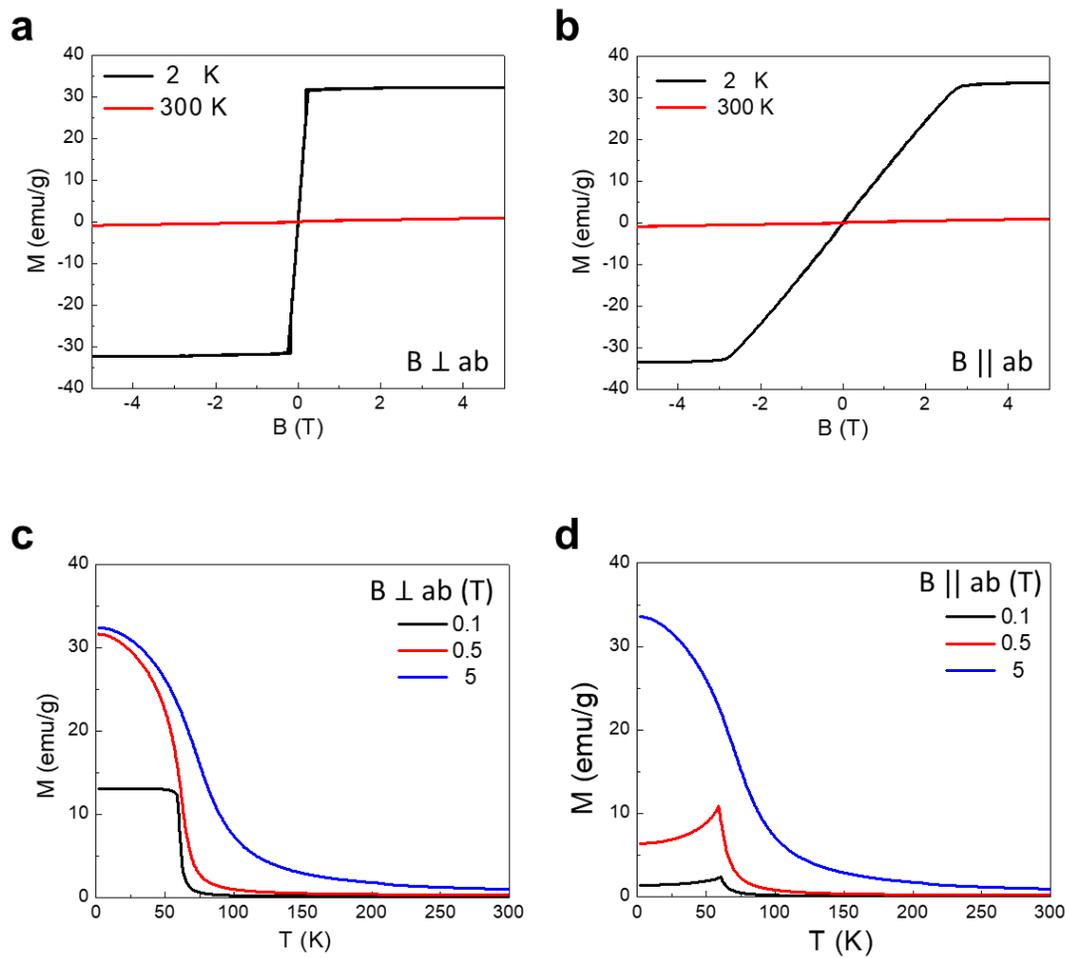

**Fig. S2.**

Magnetization of bulk crystal vs. magnetic field for (a) B⊥ab and (b) B∥ab. Temperature-dependent magnetization for (c) B⊥ab and (d) B∥ab.

## 3. Optical image of CrI₃ device

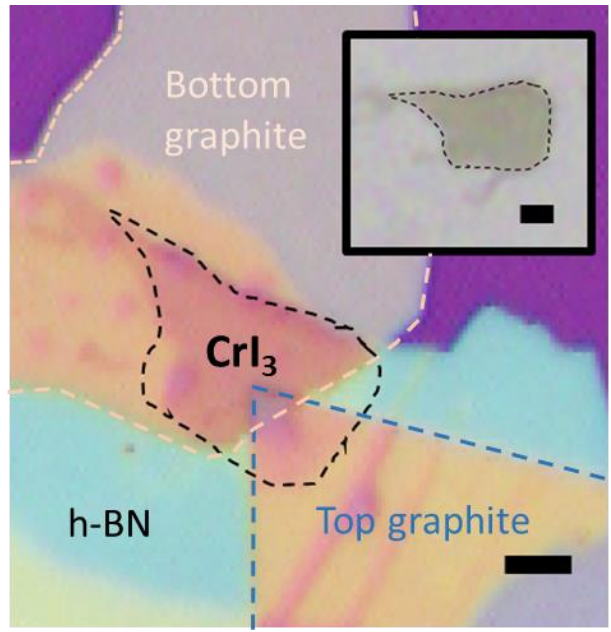

**Fig. S3**
Optical image of device with 10-layer-thick CrI$_3$ before covering with top h-BN. Inset shows transmission optical microscope image of CrI$_3$ flake. Scale bars 5 µm.

## 4. Temperature-dependent AC conductance of CrI₃ device

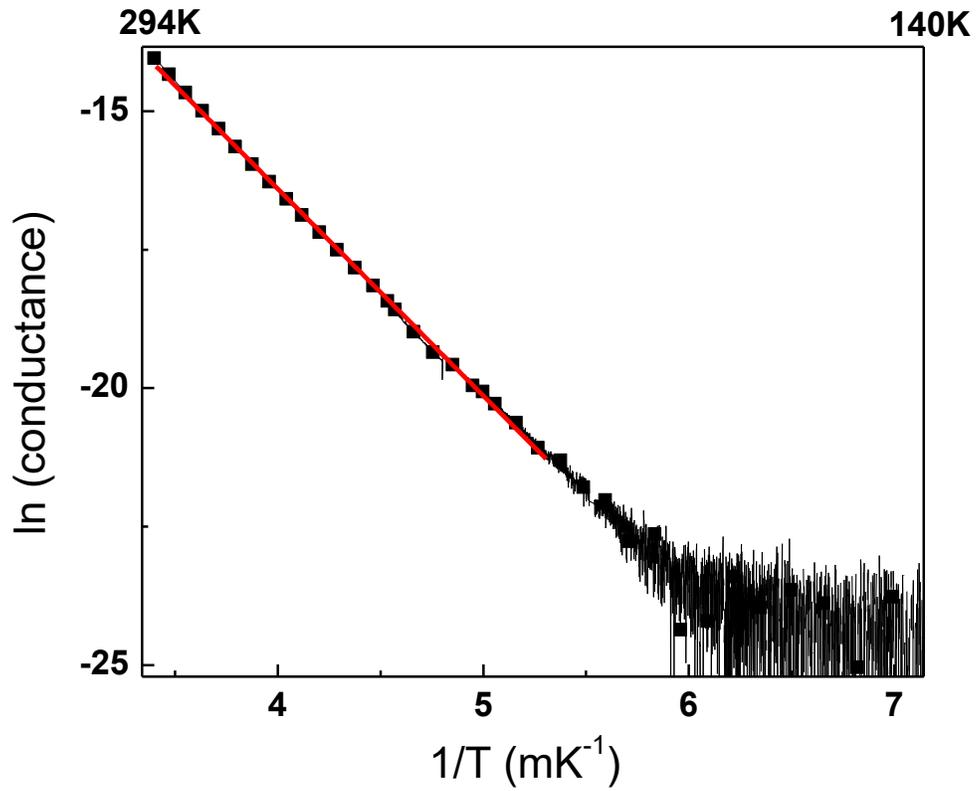

**Fig. S4**
ln(conductance) vs. 1/T of 21-layer CrI$_3$ device measured by lock-in with V$_{AC}$ = 4 mV, 17Hz excitation. The conductance follows an Arrhenius trend from 294K to 160K, with gap 322 meV.

## 5. Measurement of $I_{min}$

At the bias voltage where MR is maximum, the minimum current measured close to zero field is typically several pA in our devices, higher than both the noise and zero current level of our instrument. To check this, we measured current at several different voltages, and the results for the 8-layer device are shown below in Figure S5. At 0.31V, where MR reaches nearly $10^6$% (see Fig. S6), $I_{min}$ is well-defined at ~5.5pA.

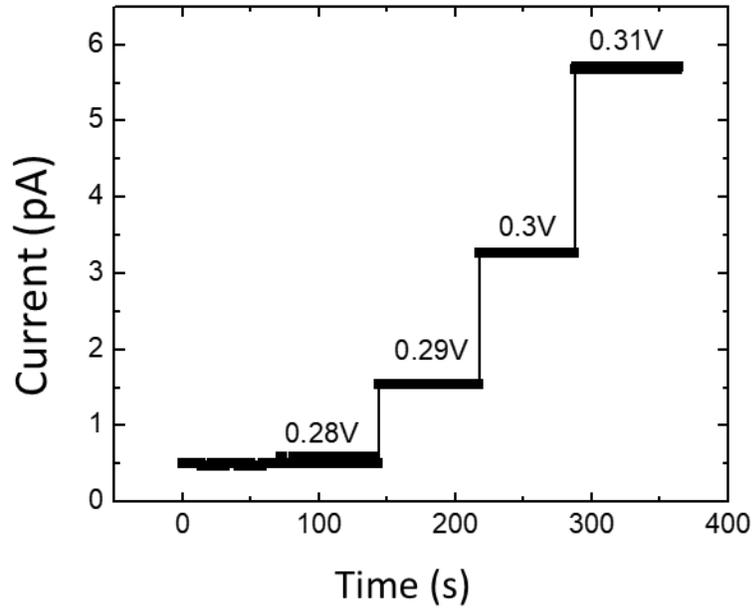

**Fig. S5**
Accuracy of small current measurements of 8-layer $CrI_3$ device at 1.4 K.

## 6. CrI$_3$ thickness dependence on (magneto)transport

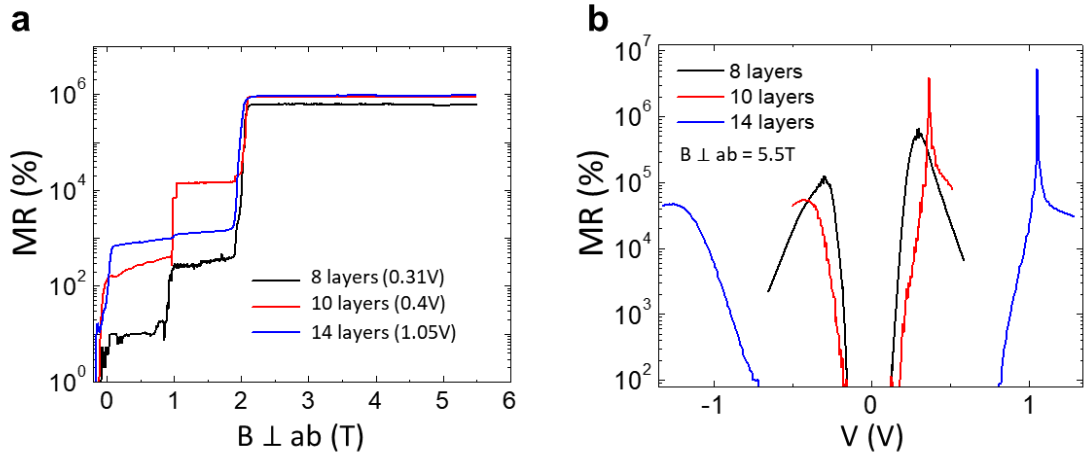

**Fig. S6**
(a) MR (%) $= \frac{I(B) - I_{min}}{I_{min}} \times 100\%$ at constant voltage vs. magnetic field for 8-, 10-, and 14-layer CrI$_3$ devices. (b) MR at 5.5T versus voltage for the three CrI$_3$ devices. The non-monotonic behavior can be understood as follows. At low bias, both *I*(5.5T) and *I*$_{min}$ are below the noise level. Increasing voltage, *I*(5.5T) increases above the noise while *I*$_{min}$ remains at the noise level, giving increasing MR. When both currents exceed the noise, MR decreases with voltage as expected for Fowler-Nordheim tunneling (see discussion in main text).

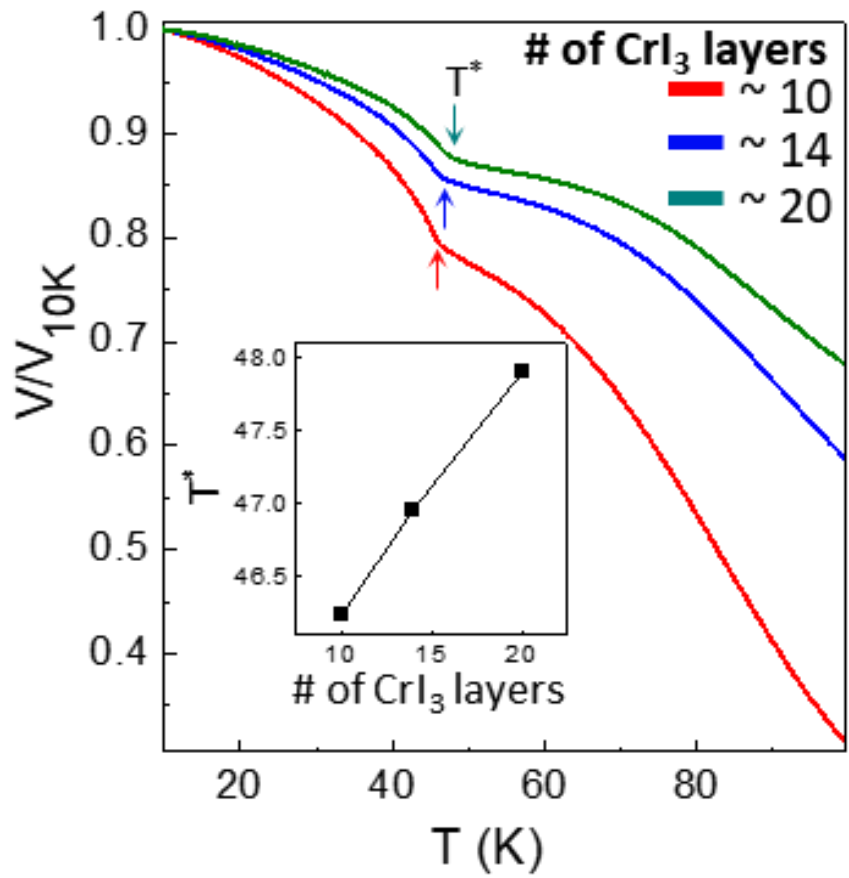

**Fig. S7**
Normalized voltage vs. temperature plot of three $CrI_3$ devices taken at zero field and 100nA current biasing. Inset shows T* vs. thickness.

## 7. Additional magnetic field sweep data

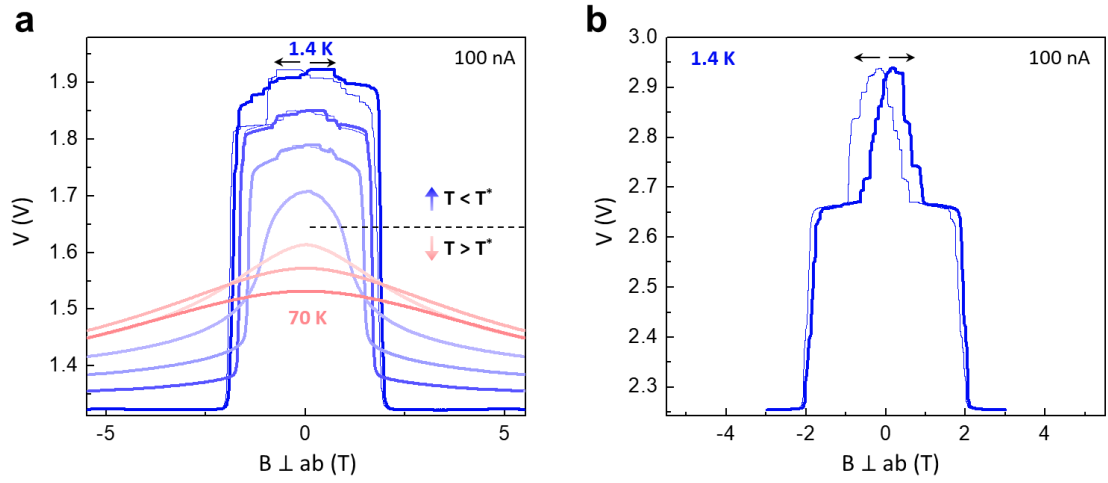

**Fig. S8**
Voltage versus magnetic field sweeps of (a) 14- and (b) 20-layer CrI$_3$ junctions.

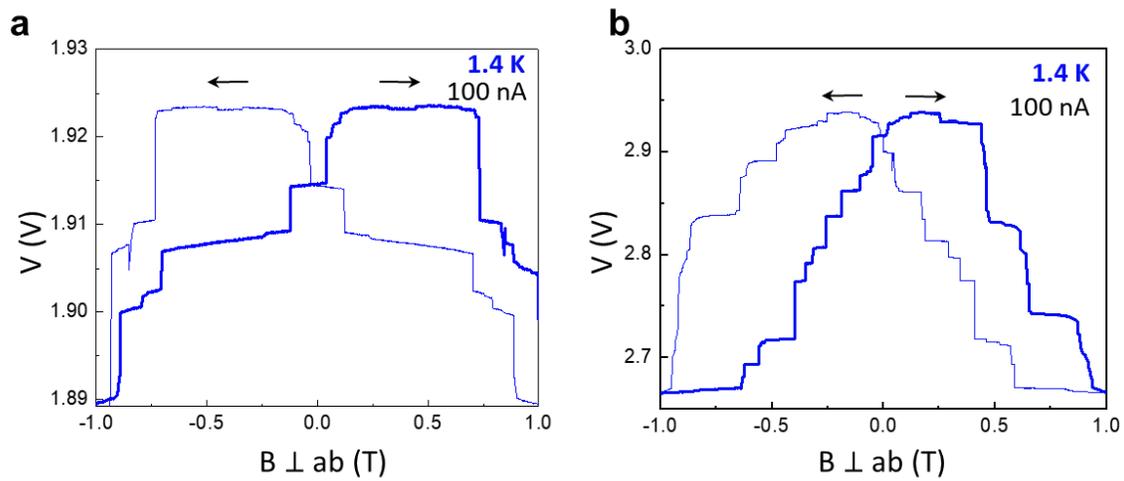

**Fig. S9**
Zoom-in plots of Fig. S8 (a) and (b) showing jumps and hysteresis.